\documentclass[12pt]{article}
\usepackage{graphicx,amssymb,amsmath}
\usepackage[colorlinks, linkcolor=red, anchorcolor=blue, citecolor=blue]{hyperref}
\newcommand{\be}{\begin{equation}}
\newcommand{\ee}{\end{equation}}
\newcommand{\bea}{\begin{eqnarray}}
\newcommand{\eea}{\end{eqnarray}}
\newcommand{\beas}{\begin{eqnarray*}}
\newcommand{\eeas}{\end{eqnarray*}}

\begin{document}
\begin{titlepage}

\vspace*{-24mm}
\rightline{LMU-ASC 05/15}
\rightline{MPP-2015-19}
\vspace{4mm}

\begin{center}

{\Large Firewalls as artefacts of inconsistent truncations of quantum geometries}

\vspace{4mm}

\renewcommand\thefootnote{\mbox{$\fnsymbol{footnote}$}}
Cristiano Germani\footnote{cristiano.germani@lmu.de}${}^{1,2,3}$ and
Debajyoti Sarkar\footnote{debajyoti.sarkar@physik.uni-muenchen.de}${}^{1,2}$
\vspace{4mm}

${}^1${\small \sl Max-Planck-Institut f\"ur Physik, F\"ohringer Ring 6, D-80805 Munich, Germany}\\
${}^2${\small \sl Arnold Sommerfeld Center, Ludwig-Maximilians-University Theresienstr. 37, 80333 M\"{u}nchen, Germany}\\
${}^3${\small \sl Institut de Ci\`{e}ncies del Cosmos, Universitat de Barcelona, Mart i Franqus 1, 08028 Barcelona, Spain}

\end{center}

\vspace{4mm}

\noindent
In this paper we argue that a firewall is simply a manifestation of an inconsistent truncation of non-perturbative effects that unitarize the semiclassical black hole. Namely, we show that a naive truncation of quantum corrections to the Hawking spectrum at order ${\cal O}(e^{-S})$, inexorably leads to a ``localised'' divergent energy density near the black hole horizon. Nevertheless, in the same approximation, a distant observer only sees a discretised spectrum and concludes that unitarity is achieved by ${\cal O}(e^{-S})$ effects. This is due to the fact that instead, the correct quantum corrections to the Hawking spectrum go like ${\cal O}( g^{tt} e^{-S})$.
Therefore, while at a distance far away from the horizon, where $g^{tt}\approx 1$, quantum corrections {\it are} perturbative, they {\it do} diverge close to the horizon, where $g^{tt}\rightarrow \infty$. Nevertheless, these ``corrections" nicely re-sum so that correlations functions are smooth at the would-be black hole horizon. Thus, we conclude that the appearance of firewalls is just a signal of the breaking of the semiclassical approximation at the Page time, even for large black holes.

\end{titlepage}
\setcounter{footnote}{0}
\renewcommand\thefootnote{\mbox{\arabic{footnote}}}

\section{Introduction}\label{sec:intro}

Within the realm of quantum field theory, it is expected that quantum effects become more and more important at smaller and smaller distances. Thus, one would infer that a large Black Hole (BH) (i.e. a BH with a radius much larger than the Planck scale $M_p$) is entirely described classically, at least, up to its own horizon. In addition, the dominant quantum corrections to the classical BH should be well described in the semiclassical limit where self-gravitational interactions are neglected, i.e. in the decoupling limit of gravity. In this context, Hawking \cite{Hawking:1974sw} showed that a BH radiates a thermal spectrum and very slowly ``evaporates".

Because of its nature, the Hawking evaporation carries no information. This, in combination with the classical no-hair theorem, stating that a BH is only described by its mass, angular momentum and charge, immediately leads to a possible (information) paradox \cite{Hawking:1976ra}:

If the Hawking evaporation continues up to a Planck size BH (the size in which semiclassical analysis breaks down), the whole information about the BH precursor (a collapsing star for example) would be lost forever and then quantum physics would be non-unitary. 

In the present form, this (information) paradox is related to the assumptions that gravity back-reactions, during the evaporation, do not dramatically change the thermal nature of the radiation\footnote{In addition, but related to that, the paradox only appears when there is a classical horizon. However this might not be the case for a collapsing star as it has been argued in \cite{no-horizon} by using holography.}. Therefore, in order to overcome the information paradox, one needs non-negligible contributions to the Hawking radiation. It has been suggested in \cite{dvali}, that the information paradox is related to the fact that quantum ${\cal O}(1/S)$ corrections (with $S$ the black hole entropy)  to the original Hawking calculation are {\it not} taken into account,  as opposed to the semiclassical ${\cal O}(e^{-S})$ ones. At the time in which the entropy is reduced to half (the Page time \cite{page}) these corrections lead to a strong departure from semiclassicality \cite{violation}. In this paper, we will show that an additional paradox arises if one carries on considering the quantum corrections to the black hole to be perturbative: the so-called firewall paradox.  It is a paradox because one would immediately conclude that the assumption of small corrections to the black hole geometry fails right at the horizon, due to a large energy stored there. 

The information paradox becomes more concrete for a big BH in AdS. In this case, a large BH semiclassical in AdS is in thermal equilibrium with its own Hawking radiation. This is because the AdS space works effectively as a box of physical size $R$ (the AdS radius)\footnote{With a BH, the radial volume of the spacetime is actually infinite. However, this is due to the infinite redshift of the horizon of the BH. Therefore, excluding the vicinity of the BH, AdS is still a box of size $R$. This is what is important for the thermodynamical equilibrium.}. However, correlation functions on this background exponentially decay after a long time \cite{barbon} making the system non-unitary. In the context of AdS/CFT \cite{Maldacena:1997re}-\cite{Witten:1998qj}, the BH configuration should be dual to a thermal CFT. A thermal CFT is however unitary. A typical signal of a unitary theory at finite temperature is a very long wavelength modulation of the correlation functions with Heisenberg time period $t_H\sim \frac{1}{T}e^{S(T)}$ \cite{barbon}, where $S$ is the entropy of the CFT and $T$ is the temperature. Smaller frequencies would instead be contaminated by thermal noise, disappearing for $T\rightarrow 0$. In addition, the unitary system is periodic at even larger time, the Poincar\'{e} time $t_P\propto e^{e^S}$ \cite{barbon}.

Because a large semiclassical BH in AdS is believed to be eternal, one might then infer that the lost information is disclosed within the BH horizon and eventually retrieved in a Heisenberg time by some $e^{- S}$ correction to the Hawking's original calculation\footnote{$e^{-S}$ is the suppression appearing for quantum fluctuations around the BH saddle point in the Euclidean path integral.}. However, as it is well understood from AdS holography, there is no way for local operators inside the BH horizon to be dual to local operator in the CFT \cite{apologia}, and thus the CFT cannot retrieve information from inside the horizon. 

We then have a clear mismatch that seemingly contradicts the AdS/CFT conjecture: there is no (unitary) CFT dual to the (non-unitary) bulk BH. The only way out for the AdS/CFT to be correct is therefore that the initial assumption that large BH can be treated semiclassically is invalid.

All information issues are related to the existence of a horizon that causally disconnects two parts of the spacetime, the exterior and the interior of the BH. Because ${\cal O}(e^{-S})$ corrections are believed to be too tiny to modify the spacetime structure of the BH (we will see that this is not true!) and no corrections are expected far away from the BH, insisting on the existence of a horizon and local physics, necessarily implies that at least part of the information must be stored within a tiny region {\it outside} the BH horizon. In this case \cite{apologia} and \cite{amps} argued that an in-falling observer would see a firewall (i.e. it would burn while crossing the horizon). Vice-versa, allowing non-local operators, that are able to retrieve information from behind the horizon would (most probably) solve the information paradox \cite{Papadodimas:2013jku,Kabat:2014kfa}. 

Note however that there is a limit in which no firewall would appear. This is the case in which the metric is classical. Here the black hole mass is infinitely large but the horizon size is finite. Classical no-hair theorem tells us that all information of the inside horizon black hole is just encoded in the black hole mass. Therefore, in this limit, the dual CFT has ``access" to the inside horizon physics, in turn, following the previous discussion, no firewalls would appear. 

The mass of a black hole in asymptotically AdS is bounded to be larger than some positive power of the cosmological constant (depending on spacetime dimensions). On the other hands, the number of fields in the dual CFT ($N$) is also proportional to the cosmological constant (in AdS/CFT). Thus, an infinite BH mass is readily obtained in the $N\rightarrow \infty$ limit. In the CFT language then, firewalls disappear in the infinitely large $N$ limit, a known result (note also that in this limit $t_H\rightarrow\infty$). A similar conclusion can also be reached from a different corner. Based on \cite{Hamilton:2005ju}-\cite{Sarkar:2014dma}, the Authors of \cite{Kabat:2014kfa, Hamilton:2007wj} noticed that the necessary non-local operators that would avoid firewalls, become local in the $N\rightarrow \infty$ limit.

An alternative resolution to the information paradox is that local quantum operators see no horizon (and thus the bulk state is described by two entangled state of CFTs\footnote{Those are the two otherwise causally disconnected AdS boundaries.}): To calculate correlation functions one should consider the integration over geometries in the path integral. What we mean here is that geometries with trivial topology, i.e. with no-horizon, would dominate the path integral over the classical saddle points at large time\footnote{Since we expect the system to be unitary, it is quasi-periodic with very very large period $t_P$. With ``large time" we mean here, and in the rest of the paper, a large time interval within this period.}, as already suggested in \cite{maldacena} and \cite{Hawking:2005kf} (see \cite{page2,forecast} for later discussions related to the firewall problem). Note that going away from the semiclassical limit, the dominant contribution to the correlation functions seems not to belong to the leading (Euclidean) saddle point approximation \cite{barbon,maldacena}; we will make this more clear later on.\footnote{An alternative approach based on stringy microstructure instead of the path integral approach is the so called fuzzball proposal \cite{Mathur:2009hf}.}

What we are going to show, is that firewalls simply arise as inconsistent expansions of non-perturbative dominant metrics (compared to other Euclidean saddle points) around a large BH background. In other words, as seen from the semiclassical BH observer, this simply means that the assumption that quantum gravity corrections to the Hawking spectrum are of order ${\cal O}(e^{-S})$ is not enough to retrieve information, unless there is a firewall at the horizon that stores it. Therefore, if the firewall is refused, quantum gravity corrections to the Hawking evaporation must be far larger than what expected, in particular, close to the horizon and after a Page time \cite{page}, they cannot even be recast as perturbative corrections to the BH metric.

\section{Black Holes and AdS/CFT at finite temperature}

As we have already mentioned, in the field theory side, one way to see the signature of unitarity is by looking at correlators. A CFT is uniquely determined by its gauge group and the dimension $N$ of matrix fields of the field theory. For a finite $N$ (but large enough to be in semiclassical approximation), the CFT spectrum is discrete. If the system is unitary and the phase space volume is compact, the correlators are periodic unless the system is dissipative (i.e. the state we are checking is not pure). In the presence of temperature, at shorter time, the system is dominated by thermal noise and the correlators do not show a periodic behaviour. However, after waiting a long time, precisely the Heisenberg time $t_H$, the system evolves into pure states and correlators start to periodically oscillate. Finally after even an exponentially longer Poincar\'{e} time $e^{e^S}$ the system approaches its original state it started with\footnote{Note the confusion of time scales definitions in \cite{solo}.}.

According to AdS/CFT, the quasi-periodic behaviour of the correlation functions is dual to the path integral calculation of the correlation functions in the gravity side.
Because of the decaying behaviour of correlation functions on a BH background, considering only the Lorentzian saddle points, would only capture the short time behaviour of the correlation functions in the CFT. In two dimensions for example,
in \cite{Germani:2013sra}, it has been shown that, in the semiclassical limit, Euclidean saddle points are as dominant as Lorentzian's.
In three dimensions the same conclusion was shown in \cite{maldacena}:

Considering, as in the rest of the paper, a Ba\~{n}ados, Teitelboim and Zanelli (BTZ) black hole \cite{Banados:1992wn}, Maldacena showed that the correlation functions calculated on a thermal AdS (TAdS) (an Euclidean saddle point), becomes as large as the ones calculated on a BH's background, after large time \cite{maldacena}. Specifically:
\begin{equation}\label{mal_paper}
\langle\mathcal{O}(x)\mathcal{O}(0)\rangle\approx e^{-S_{BTZ}}\langle\mathcal{O}\mathcal{O}\rangle_{\mbox{BTZ}}+e^{-S_{TAdS}}\langle\mathcal{O}\mathcal{O}\rangle_{\mbox{TAdS}}
\end{equation}
where $S_{BTZ}$ and $S_{TAdS}$ are the Euclidean action of BTZ and TAdS respectively. Nevertheless, although the TAdS contributes to the large time behaviour of the correlation functions, it still does not help to retrieve the correct Heisenberg time for the finite $N$ case. It was then argued by Maldacena \cite{maldacena} that one should therefore consider all $SL(2,Z)$ family of the BTZ black holes in the calculation of the correlation functions.

It was later shown in  \cite{solo} and \cite{Kleban:2004rx} that, unfortunately, even the sum over these topological diversities does not reconcile with the CFT behaviour at large time. Furthermore, the Authors in \cite{Kleban:2004rx} showed that at the time scale $t_0$ in which finite entropy effects becomes important, the perturbation theory on gravitational saddle points necessarily break down. After this time, the correlation functions must then be dominated by the next-to-leading order approximation in saddle points: usually, in the semiclassical limit of quantum gravity, a path integral is well approximated by the leading saddle point approximation, i.e., given some operator $\cal O$ of metric and matter fields $\Phi$ we have (after Wick rotation)
\begin{eqnarray}
\langle\mathcal{O}(x)\mathcal{O}(0)\rangle &=&N\int {\cal D}\Phi{\cal D}g {\cal O}(\Phi(x),g(x)){\cal O}(\Phi(0),g(0)) e^{-S(\Phi,g)}\simeq \cr
N &\sum_{g=g_i}& \int {\cal D}\Phi \left[{\cal O}(\Phi(x),g_i(x)){\cal O}(\Phi(0),g_i(0)) +\cdots\right]e^{-S(\Phi,g_i)}\ ,
\end{eqnarray} 
where $g_i$ are the gravitational saddle points obtained by solving $\delta S/\delta g=0$, and $N$ is a normalisation factor. The ellipses denotes the next to leading order terms in the saddle point approximation and are proportional to subsequent derivatives of the operators $\cal O$ with respect to the metric, e.g. the first of those is proportional to $\frac{\delta^2 \left[{\cal O}(\Phi(x),g(x)){\cal O}(\Phi(0),g(0))\right]}{\delta g^2}\Big|_{g=g_i}$. Since, as we already noticed, the correlation functions on a BH background decay to zero at large time, the true correlations functions must be dominated by the next-to-leading order approximation to the saddle point.
The evaluation of these integrals is obviously an almost impossible task and the saddle point approximation ceases to be helpful. It is then a common practice to hope for the existence of some localized geometries in the path integral that keep a geometrical description of the BH correlation functions. Obviously this is an assumption that has to be eventually proven, following the literature on the topic, here we use it as a working assumption. In any case, even if this assumption turns out to be incorrect, our findings are the proof of the concept that firewalls only appear as a naive ${\cal O}(e^{-S_{\rm BH}})$ expansion of the path integral of gravity.

Following the  Hawking conjecture that only trivial topologies dominate on long time scales \cite{Hawking:2005kf} we can then try to search for a topologically trivial metric $g_{\rm NP}$ such that
\begin{equation}\label{boh}
\langle\mathcal{O}(x)\mathcal{O}(0)\rangle\Big|_{t\gg t_0}\approx \langle\mathcal{O}(x)\mathcal{O}(0)\rangle\Big|_{g=g_{\rm NP}} e^{-S_{NP}} \ .
\end{equation}
This question has been answered by Solodukhin in \cite{solo}. The Author proposed a candidate smooth metric that is subdominant at early time and then dominates over BTZ and TAdS at $t\gg t_0$ so as to reproduce the expected long wave oscillation of the correlation functions with period $t_H$. In this paper, we consider (\ref{boh}) and show that firewalls emerge in (unitary) correlation functions whenever the metric expansion $g_{\rm NP}=g_{\rm BTZ}+{\cal O}(e^{-2S})$, is performed\footnote{Note that, in order to have a ${\cal O}(e^{-S})$ correction to the Hawking spectrum the metric must be corrected at order ${\cal O}(e^{-2S})$.}. In fact the situation is even more dramatic: If (\ref{boh}) is valid, there is no parameter, say $\alpha$, such that $g_{\rm NP}=g_{\rm BTZ}+{\cal O}(\alpha)$ and firewalls are avoided. On the contrary, by using the full non-perturbative metric no firewall appear. 

Note that the firewall resolution of \cite{er-epr}, the so called ``ER=EPR" proposal, can be reinterpreted as a close parent of the proposal of \cite{solo} aiming for a resolution of the unitary problem for BTZ black holes. Both resolutions can indeed be naively understood as the existence of a quantum wormhole connecting the two timelike asymptotic regions of the BTZ solution. The main difference be that the resolution of \cite{solo} involves a traversable wormhole, i.e. a non-semiclassical contribution. We fail to see how a non-traversable wormhole as in \cite{er-epr} could solve the unitarity problem of the BTZ. If this was true then semiclassical corrections would be enough for unitarity, a conclusion contradicting \cite{maldacena}.

\section{A warm up: the would-be firewall in Rindler}

\begin{figure}
\begin{center}
\includegraphics[width=0.6\textwidth, height=0.25\textheight]{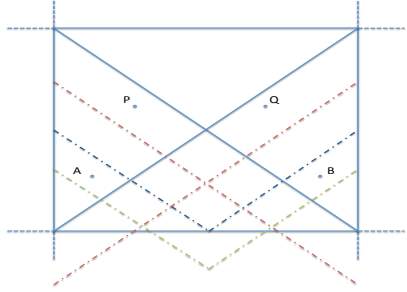}
\end{center}
\caption{Rindler patches of flat space for different accelerated observers (labeled by their horizon's color). Even though for the blue observer the points $P$ and $Q$ are non-entangled (in this case, non-causal), they are entangled for the red, purple and green observer. Similarly, points $A$ and $B$ are non-entangled for blue, red and purple observers, but entangled for green observer.}\label{fig_3}
\end{figure}

In this section, following \cite{Czech:2012be}, we will review the fact that a Rindler observer would be tempted to declare that at its own horizon there should be a firewall. Of course, we already know that this cannot be possibly true as the Rindler space is just a section of a Minkowski space. The reason the Rindler observer would declare that there is a firewall is due to the fact that this observer has no access to the physics behind its own horizon and would naively think that quantum operators have the same blindness. However, a judicious path integral calculation that includes the Minkowski metric, would show that the information behind the horizon is in fact retrievable. Of course, in this case, the Rindler space is just a coordinate transformation from the Cartesian coordinates, but for the Rindler observer, the  Minkowski space is a new smooth  geometry that has to be added to the path integral calculation. The Rindler observer will then understand the fact that in its own space, there are always observers that can see behind its own horizon. Therefore, he/she would understand that quantum operators in the left and right Rindler wedges are in fact entangled, as well as those inside and outside the Rindler's horizon (see figure \ref{fig_3}). In other words he/she would quantum mechanically discover diffeomorphisms! 

\begin{figure}
\begin{center}
\includegraphics[width=0.6\textwidth, height=0.25\textheight]{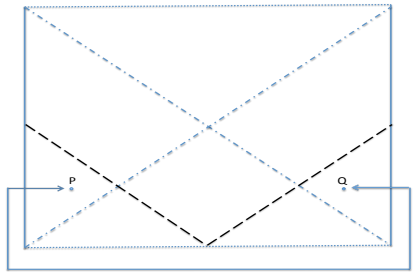}
\end{center}
\caption{Eternal or BTZ BH spacetimes. The black dashed lines are the light cones of the observer for whom most of the spacetime outside the horizon is causally connected. However, still e.g. points like $P$ and $Q$ are causally disconnected to her at finite $N$.\label{fig}}
\end{figure}

Conversely to the trivial topology case (e.g. Rindler), because of the end of space (the singularity) in non-trivial topologies (e.g. a BH), there are no observers who can entangle operators in the whole spacetime (see figure \ref{fig}). 

\subsection{Rindler firewall: a mini review}

The appearance of a firewall for a Rindler observer has been proven in the case of compact Rindler space \cite{Czech:2012be}. 

\begin{figure}
\begin{center}
\includegraphics[width=0.6\textwidth, height=0.3\textheight]{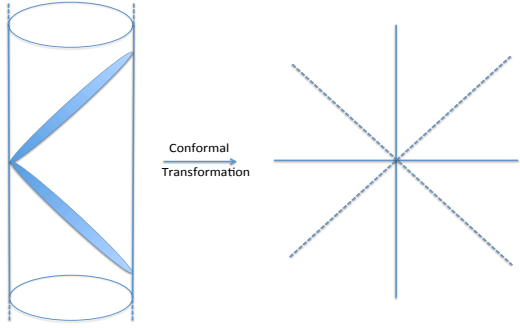}
\end{center}
\caption{Conformal transformation from Poincar\'{e} (the right triangular region of the global AdS cylinder bounded by lightcones) to Minkowski spacetimes. One can do another conformal transformation to go to the hyperbolic CFT's.\label{fig_2}}
\end{figure}

In \cite{Czech:2012be} it was argued that pure AdS spacetimes have a dual description in terms of an entangled pair of two hyperbolic space CFT's which can be seen by conformal transforming the Poincar\'{e} patch of AdS to Minkowski (and then to hyperbolic spaces) (see figure \ref{fig_2}), which are related to Rindler spaces with metric
\be
ds^2_{\rm Rind}=-a^2 y^2dt^2+dy^2\ .
\ee
These hyperbolic spaces, related to an accelerated observer with acceleration $a$, see an Unruh temperature $T=\frac{a}{2\pi}$ \cite{unruh} (we use here and for the rest of the paper units $\hbar=c=1$). In Euclidean time, this implies that typical quantum states in this space are thermal and get a Boltzmann factor $e^{-\frac{\beta E}{2}}$ ($\beta\equiv T^{-1}$): 
\begin{align}
|0_{{\rm global}\ AdS}\rangle=|0_{S^d}\rangle\to|0_{Mink}\rangle&=\frac{1}{Z}\sum_{i}e^{-\frac{\beta E_i}{2}}|E_{i(Rind)}^L\rangle\otimes|E_{i(Rind)}^R\rangle\nonumber\\
&=\frac{1}{Z}\sum_{i}e^{-\pi R_H E_i}|E_{i}^L\rangle_{H^d}\otimes|E_{i}^R\rangle_{H^d}
\end{align}
with $R_H$ being the radius of the hyperbolic spacetimes, $Z$ is the partition function and $E_i$ are the energy levels. This is just like the entangled description of Minkowski space in terms of two Rindler patches (or the description of eternal Schwarzschild BHs in AdS in terms of entanglement between two CFT states \cite{maldacena}).

The Authors in \cite{Czech:2012be} considered a mass-less scalar field $\phi$ in a two dimensional Minkowski spacetime (with coordinates $T$ and $Z$) and expanded the field in Rindler modes (we focus for simplicity solely to right handed modes)
\[
\phi=\int_{0}^{\infty}d\omega(b_{\omega,R}\phi_{\omega,R}+b^\dagger_{\omega,R}\phi^*_{\omega,R})+L\leftrightarrow R\ .
\]
where, with $U=T-Z$ and $V=T+Z$,
\begin{align}\label{modes}
&\phi_{\omega,R}(U)=\Theta(-U)\frac{1}{\sqrt{4\pi\omega}}(-aU)^{i\omega/a}\quad\mbox{and}\nonumber\\
&\phi_{\omega,L}(U)=\Theta(U)\frac{1}{\sqrt{4\pi\omega}}(aU)^{-i\omega/a}\ .
\end{align}
The energy momentum tensor for the massless scalar field is
\be
T_{\mu\nu}=\partial_\mu\phi\partial_\nu\phi-\frac{1}{2}(\partial \phi)^2g_{\mu\nu}\ .
\ee
It is then clear that upon using the Rindler modes, the stress tensor has a pole in $U=0$ corresponding to the horizon. One should then regularise the modes expansions \eqref{modes} in order to use them in the energy momentum tensor \cite{parentani}. The regularised mode expansions are 
\begin{align*}
&\phi_{\omega,R}(U)=\frac{1}{\sqrt{4\pi\omega}}\frac{(a(U-i\epsilon))^{i\omega/a}-(a(U+i\epsilon))^{i\omega/a}}{e^{\pi\omega/a}-e^{-\pi\omega/a}}\quad\mbox{and}\\
&\phi_{\omega,L}(U)=\frac{1}{\sqrt{4\pi\omega}}\frac{e^{\pi\omega/2}(a(U-i\epsilon))^{-i\omega/a}-e^{-\pi\omega/2}(a(U+i\epsilon))^{-i\omega/a}}{e^{\pi\omega/a}-e^{-\pi\omega/a}}\ ,
\end{align*}
where $\epsilon$ is an arbitrarily small parameter.

Focusing on the $UU$ component, we have
\begin{eqnarray}\label{tobe}
\langle T_{UU}\rangle=\langle 0,L|\langle 0,R|(\partial_U\phi(U))^2 |0,R\rangle|0,L\rangle-\langle 0,Mink|(\partial_U\phi(U))^2|0,Mink\rangle\nonumber\\
=-2\int_{0}^{\infty}d\omega[\beta_\omega^2(|\partial_U\phi_{\omega,R}|^2+|\partial_U\phi_{\omega,L}|^2)+2\alpha_\omega\beta_{\omega}\mbox{Re}(\partial_U\phi_{\omega,L}\partial_U\phi_{\omega,R})]\ ,
\end{eqnarray}
with $\alpha_\omega=e^{\pi\omega/a}\beta_\omega=(1-e^{-\pi\omega})^{-1/2}$ being the Bogliubov coefficients connecting the Minkowski with the Rindler vacuum. In the above, the UV renormalisation of the stress tensor is obtained by subtracting the Minkowski counterpart and by using the transformations between the Minkowski and Rindler vacuums 
\[
|0,Mink\rangle=\mathcal{U}|0,Rind\rangle \quad\mbox{with}\quad \mathcal{U}=\prod_{\omega>0}\exp (-\tan^{-1}(e^{-\pi\omega/a}))(b^\dagger_{\omega,R}b^\dagger_{\omega,L}-b_{\omega,R}b_{\omega,L})\ .
\]
It is now a tedious but straightforward computation to find \cite{parentani}
\be\label{div1}
\langle T_{UU}\rangle\approx - \frac{1}{\epsilon^2}\ .
\ee
This divergence is precisely the hypothetical firewall that a Rindler observer would calculate. In terms of the energy density ${\cal E}\equiv \langle T^t{}_t\rangle$ as seen by a Rindler observer, it is easy to see that
\be
{\cal E}\propto - \frac{e^{2 a t}}{\epsilon^2}\ .
\ee

Note that this divergence can never be canceled out if one considers the stress tensor in a general product state
\begin{align*}
|\psi\rangle=&\sum_{k,n_i=1}^{\infty}\int\left(\prod_{i=1}^{k}d\omega_i\right)f_{(n_1,\dots,n_k)}(\omega_1,\dots,\omega_k)\left(\prod_{i=1}^{k}(b^\dagger_{\omega_i,L})^{n_i}\right)|0\rangle_L\nonumber\\
&\otimes\sum_{k,n_i=1}^{\infty}\int\left(\prod_{i=1}^{k}d\omega_i\right)g_{(n_1,\dots,n_k)}(\omega_1,\dots,\omega_k)\left(\prod_{i=1}^{k}(b^\dagger_{\omega_i,R})^{n_i}\right)|0\rangle_R\ .
\end{align*}
This is due to separability of the states and to the fact that each term contributes to $T_{UU}$ non-negatively\footnote{Here $b,b^\dagger$'s are annihilation-creation operator corresponding to the scalar field, $\omega$'s are conjugate momenta and $f$ and $g$ are general momentum dependent coefficients of the product state. It can be shown that each terms of the $T_{UU}^\psi$ are either complete squares or are rapidly oscillating function of momentum around $U\to 0$, so that upon integrating them over momentum gives zero.}. In other words, without entangled states between the right and left side of the Rindler wedge, the dangerous divergence of the stress tensor cannot be removed.

On the other hand, for an entangled state
\[
|\psi_1\rangle=\int d\omega f(\omega)b^\dagger_{\omega,L}|0\rangle_{L}|0\rangle_{R}
\]
with 
\[
f(\omega)=\frac{e^{-\frac{(\omega-\omega_0)^2}{2\delta^2}}}{\sqrt{2\pi}\delta}
\]
one has, for $aU\in[e^{-a/\delta},e^{a/\delta}]$,
\be
\langle T_{UU}^{\psi_1}\rangle=2\Big|\int d\omega f(\omega)\partial_U\varphi_{L\omega}\Big|^2=\frac{\omega_0^2}{2\pi a^2 U^2}e^{-\frac{\delta^2(\log a U)^2}{a^2}}\ .\label{psi1}
\ee
Taking $\delta=-\frac{a}{\ln a\epsilon}$, the stress tensor \eqref{psi1} can be evaluated at the regularised horizon $U=\epsilon$. One then finds 
\be
\langle T_{UU}^{\psi_1}\rangle\Big|_{U=\epsilon}\approx +\frac{1}{\epsilon^2}\ ,
\ee
which is precisely the same divergence \eqref{div1} but with opposite sign. What we have discovered here is that, by an appropriate choice of $\omega_0$, the divergence of the stress tensor calculated from the point of view of a Rindler observer is precisely canceled out by the divergence of the stress tensor constructed by using $L$ and $R$ entangled states defined by $\psi_1$. Since the left and right wedges of the Rindler space {\it are} entangled from the point of view of a non-accelerated observer, we are in fact allowed to use the $\psi_1$ state. Thus, this proves the non-existence of a firewall, despite what a Rindler observer would guess.

On a BH spacetime, instead, as discussed before, no entangled state connecting the two future asymptotic regions, separated by the horizon, exist. 

\section{Firewall in BTZ}

We will now consider a non-rotating BTZ black hole metric in AdS$_3$ which is given by \cite{Banados:1992wn}
\begin{equation}\label{ads_BTZ}
ds^2=\left(8GM-\frac{r^2}{R^2}\right)dt^2+\left(-8GM+\frac{r^2}{R^2}\right)^{-1}dr^2+r^2d\phi^2\ .
\end{equation}
Here $M$ is the ADM mass, $R$ is the AdS radius and $G$ is 3-dimensional Newton's constant which has dimension of length. The BTZ horizon ($r_H$), temperature ($T_{BTZ}$) and entropy ($S$) are (see e.g. \cite{Cadoni:2009tk})
\begin{eqnarray}
r_H=R\sqrt{8GM}\ ,\ T_{BTZ}=\frac{\sqrt{8GM}}{2\pi R}\ ,\ S=\frac{\pi r_{H}}{2 G}\ .
\end{eqnarray}
For $r_H\geq R$ the black hole has a large but finite entropy and is in thermal equilibrium. Thus, without considering any quantum corrections, it is a so-called eternal black hole, or big black hole in AdS. This is the case we shall consider. 

By using the coordinate transformation $r=\sqrt{8GM}R\cosh \frac{y}{R}$, that map the BH exterior to real $y$ and the interior to complex $y$, we have
\begin{equation}\label{nonrotbtz}
ds^2=-8GM\sinh^2\frac{y}{R}dt^2+dy^2+8GMR^2\cosh^2\frac{y}{R}d\theta^2\ .
\end{equation}
In these coordinates, the horizon is at $y_H=0$. The near horizon geometry is of Rindler type. Suppressing the angular direction, we have for $y\sim 0$
\begin{equation}\label{nh}
ds^2_{\rm n.h.}=-a_{\rm BH}^2 y^2dt^2+dy^2\ ,
\end{equation}
which is a Rindler metric with $a_{\rm BH}=\frac{\sqrt{8GM}}{R}$.

We can now define the light-cone coordinates $U,V$ by performing the following coordinate transformations 
\[
T=y\sinh\left(\frac{\sqrt{8GM}}{R}t\right)\quad\mbox{and}\quad X=y\cosh\left(\frac{\sqrt{8GM}}{R}t\right)
\]
and then $U=T+X$ and $V=X-T$. Then we have 
\begin{equation}\label{rinddiv_2}
U= e^{\sqrt{8GM}\frac{t}{R}} y\ .
\end{equation}
Thus, the future horizon is at $U=0$. In complete analogy to the Rindler case we then find that the energy density close to the horizon diverges as  
\be
{\cal E}\propto - \frac{e^{2a_{BH}t}}{\epsilon^2}\ .
\ee
In the BTZ case, the states living inside and outside the horizon are not entangled (see figure \ref{fig}). Thus this divergence cannot be cancelled out by entangled states. However, is this divergence really there?

Let us make a quick digression. Minkowski space is not a perturbation of Rindler space. The reason is that Rindler space is just a subsection of Minkowski. Therefore, although the two spaces are trivially related to each other by diffeomorphisms (or just by analytical extensions), we could have not removed the firewall arising in Rindler by ``quantum correcting" the Rindler space. Note however, contrary to the black hole case, that the field theory on the Rindler wedge is {\it unitary}. This is due to the fact that Rindler space has an infinite entropy and thus, all information is encoded in the classical parameters, the acceleration of the Rindler observer.

Conversely to the Rindler case, for a finite size BH, where the entropy $S$ is finite, correlation functions of {\it semiclassical} solutions decay in time \cite{barbon}. This is the essence of the information paradox for BH. However, if the whole information is retrieved by quantum corrections of order ${\cal O}(e^{-S})$ to the hawking spectrum (going to zero at infinite entropy) they can never entangle states from the inside to the outside of the BH. Thus, although these quantum back-reactions to the Hawking evaporation may solve the information paradox, they cannot solve the firewall paradox, i.e. a divergent {\it negative} energy-density at the horizon similar to the one we found before. This is the message of \cite{amps}.  

More clearly: {\it Resolving the BH information by ${\cal O}(e^{-2S})$ corrections to the classical metric immediately implies a firewall at the black hole horizon}.

In the following we are going to show that this is indeed true but also that the expansion of quantum corrections in terms of ${\cal O}(e^{-S})$ breaks down precisely as soon as the firewall emerges. Specifically, we will show that ``corrections" to the classical metric go instead like ${\cal O}(e^{-S} g^{tt})$. These corrections are small far away but diverge at the BH horizon where $g^{tt}\rightarrow \infty$. Nevertheless, they re-sum to a smooth non-perturbative geometry without firewalls. In other words, any truncation of non-perturbative quantum effects at order ${\cal O}(e^{-S} g^{tt})$, even solving the information paradox, would inexorably generate a firewall. Thus:

{\it The firewall is just a mirage due to an inconsistent truncation of non-perturbative corrections to the classical metric}.

\section{The fake firewall}

Usually, in quantum field theory classical trajectory dominates the path integral. However, sometimes, non-classical configuration may be as important as the classical one, e.g. QCD instantons. 

In the BTZ case, correlation functions calculated on the classical background exponential decay in time, as already said many times. Nevertheless, they dominate the path integral at early time. Thus, at early time, the BTZ geometry dominates over all the other paths in the gravity path integral.

At later time, however, when the correlation functions calculated on the BTZ geometry decay, the off-shell configurations become more important. One may try with Euclidean saddle points, such as thermal AdS, however, as discussed already, this contribution does not retrieve unitarity so one needs to go beyond the Euclidean saddle point approximation. Assuming that at late time one can select a specific metric dominating the correlations, Solodukhin proposed a candidate metric that would match the very long time $t\sim t_H$ behaviour of the CFT correlation functions. As we mentioned already, this metric should be smooth in the sense that the whole spacetime should be entangled. With this in mind the proposed non perturbative metric $g^{\tiny NP}_{\mu\nu}$ is \cite{solo}
\begin{equation}\label{nonpertbtz_2}
ds^2=-(8GM\sinh^2(y/R)+\lambda^2)dt^2+dy^2+8GMR^2\cosh^2(y/R)d\theta^2\ .
\end{equation}
Here $\lambda$ is a non-perturbative, small, correction. Note that this metric rapidly approaches the BTZ metric away from the would be horizon and therefore it has the same AdS asymptotic as BTZ. We could then use, also in this case, AdS/CFT. 

Without $\lambda$, i.e. in the BTZ background, there are no normalisable modes for a scalar field. This is due to the fact that the effective potential exponentially decays to zero at the horizon. The factor $\lambda$ however produces a potential barrier before the horizon. This potential barrier allows for normalisable modes. The minimal frequency, corresponding to the largest time scale of the system, i.e. the expected Heisenberg time, is proportional to $\lambda$. One then fixes 
\be\label{lambdaconvention}
\lambda=\frac{\sqrt{16 G M}}{2\pi}e^{-S}\ .
\ee
Thus, as it should, for an infinitely entropic black hole this metric becomes BTZ. Note that in order to have the correct correction to the Hawking spectrum of order $e^{-S}$, and so to obtain the correct Heisenberg time, the metric has to be corrected by order $e^{-2S}$. This is clear as the metric is associated to the square of frequencies.

Let us pause here for a moment. It would seem that the non-perturbative metric (\ref{nonpertbtz_2}) is just an ${\cal O}(e^{-2S})$ correction to the BTZ BH metric, as one naively expects from a semiclassical treatment. However, although this is true far away from the horizon, it is not so, close to the horizon. This is due to the fact that, even though tiny, $\lambda$ completely changes the topology of the manifold. It is precisely for this reason that insisting on treating the metric (\ref{nonpertbtz_2}) as a perturbation of the BTZ BH would require a firewall at the horizon, as we are going to show. In this sense the firewall will just signal the breaking of the perturbation treatment. Note that, although the equivalence principle is violated by the contribution of \eqref{nonpertbtz_2}, correlations functions are smooth everywhere and the stress tensor never explodes. In other words, what we are concerned about is whether the near horizon geometry is described by a smooth wavefunction. Specifically in this sense, we then conclude that the geometry of \cite{solo} does not produce a firewall. Indeed, the firewall is described by a divergent stress tensor, which obviously also violate the equivalence principle.

One question that remains to be answered is when the non-perturbative metric (\ref{nonpertbtz_2}) would take over the BTZ metric for the calculation of correlation functions.   

In \cite{amps}, for an evaporating black hole, the Authors have suggested that the firewall appears after the Page time \cite{page} in which the entropy of the system is half of what we started from. As proven by Page, after this time, the system, if unitary, as it is assumed to be a black hole, starts to release information.

At the semiclassical level our black hole is eternal, meaning that the black hole is in thermal equilibrium with its own Hawking temperature. However, if this was true at the full quantum level, there would be a contradiction with the assumption of unitarity, as Hawking process is not unitary. Therefore,  even if big, an ``eternal" black hole must  release information breaking the thermal equilibrium. Each bit of information released, even if adiabatically as we are assuming, will then diminish the entropy of the black hole.
As discussed in \cite{apologia}, this re-introduced the concept of Page time even for big asymptotically AdS black holes. We then expect that the time $t_c$ in which the non-perturbative corrections start to dominate is precisely the Page time. 
  
\subsection{Firewall as a $e^{-S}$ expansion}

As discussed before, at very long time $t\gg t_0$, any correlation functions in the path integral must be dominated by non-saddle points of the gravitational action. In particular, following \cite{solo}, we will assume that the correlators will be dominated by the metric (\ref{nonpertbtz_2}). 

We consider a massless test scalar field $\phi$ with Lagrangian
\be\label{lag}
\mathcal{L}_{NP}=-\frac{1}{2}\sqrt{-g_{NP}}g^{\alpha\beta}_{NP}\partial_\alpha\phi\partial_\beta\phi\ .
\ee
Taking the point of view of a BTZ observer, with the expansion around the saddle point in mind, he/she would state that the quantum contributions to the metric are of order ${\cal O}(e^{-2S})$, leading to corrections of  ${\cal O}(e^{-S})$ to the Hawking spectrum.  

Thus, keeping only order $\lambda^2$ in \eqref{lag} we have
\be\label{lag2}
\mathcal{L}_{\lambda^2}=-\frac{1}{2}\sqrt{-g}\left[g^{tt}\left(1-\frac{\lambda^2}{2} g^{tt}\right)\dot\phi^2+g^{ij}\left(1+\frac{\lambda^2}{2} g^{tt}\right)\partial_i\phi\partial_j\phi\right]\ ,
\ee
where the metric $g_{\alpha\beta}$ is the BTZ metric and $\dot{}\equiv\partial_t$. Since we are considering the expansion of the non-perturbative metric as a background-dependent ``quantum correction" to the BTZ metric, we should not be surprised that the action \eqref{lag2} breaks Lorentz invariance, as does the background. 

We will now analyse the theory \eqref{lag2}. Firstly we notice that the canonically normalised field (in coordinates $(y,t)$)
\be
\varphi_y=\left[\sqrt{-g}g^{tt}\left(1-\frac{\lambda^2}{2} g^{tt}\right)\right]^{1/2}\phi
\ee
is strongly coupled when the normalisation factor 
\be
{\rm Norm}=\left[\sqrt{-g}g^{tt}\left(1-\frac{\lambda^2}{2} g^{tt}\right)\right]^{1/2}
\ee
goes to zero. This happens for $\frac{\lambda^2}{2} g^{tt}=1$, i.e. at (taking small $y$ approximations)
\be 
y_c\simeq e^{-S} \frac{R}{2\pi}\ ,
\ee
where we have considered the approximation of large entropy. As it is expected, the strong coupling effect is removed in the infinite entropy limit as there $y_c$ coincides with the position of the horizon $y_H=0$.

Schematically, the theory \eqref{lag2} is of the following type\footnote{Without lost of generality, we have integrated out the angular direction and assumed the $\phi$ field is independent of $\theta$.}
\be
A=\frac{1}{2}\int dy dt\left[-\alpha(y)^2\dot\phi^2+\beta(y)^2(\partial_y\phi)^2\right]\ ,
\ee
where $\alpha$ and $\beta$ can be easily read off from \eqref{lag2}. 

\begin{figure}
\begin{center}
\includegraphics{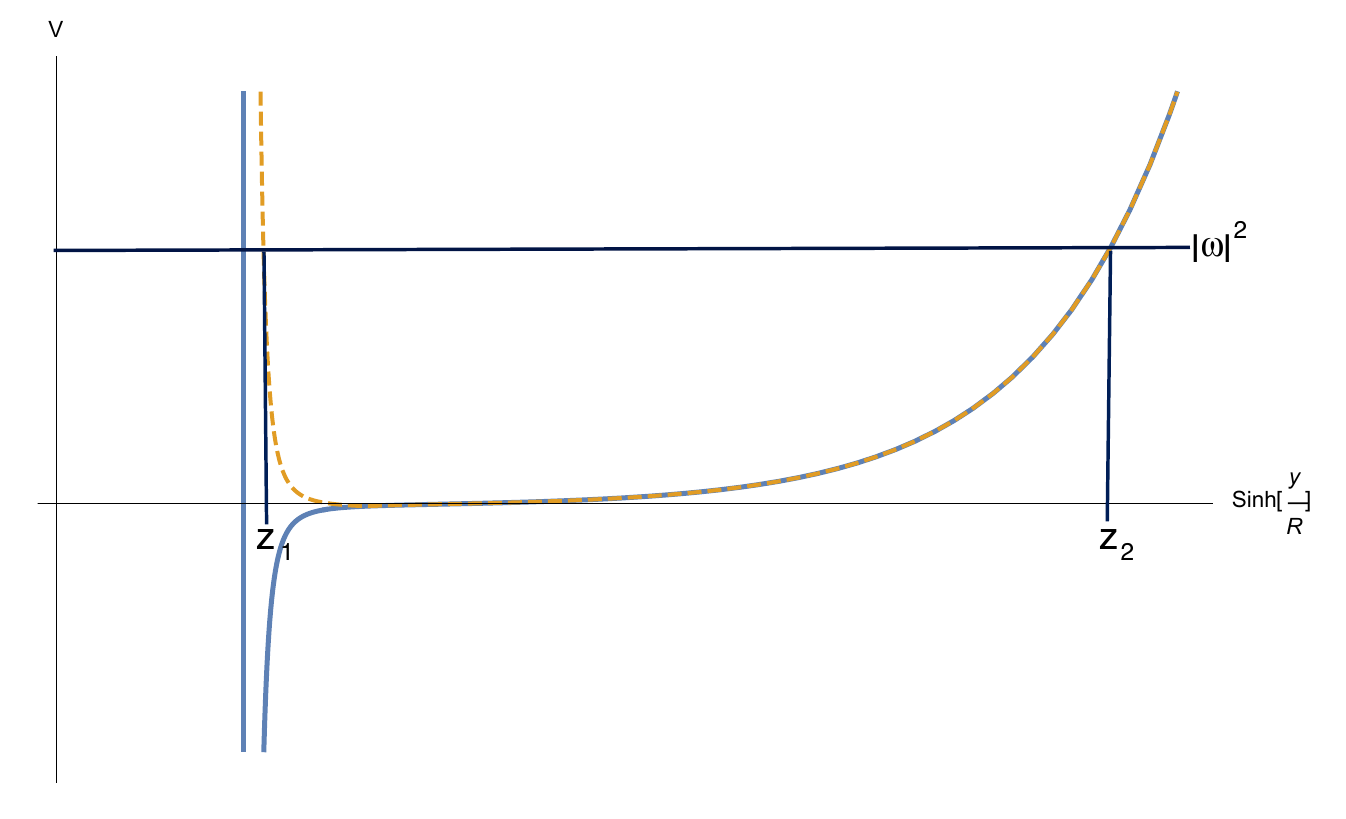}
\end{center}
\caption{The thick line is effective potential $V(z(y))$ while the dashed line is $|V(z(y))|$.\label{divV}}
\end{figure}

To handle the system, it is easier to define a new radial variable
\be
z=\int \frac{\alpha}{\beta} dy\ ,
\ee
so that the action now reads
\be
A=\frac{1}{2}\int dz dt\frac{\alpha^3}{\beta}\left[-\dot\phi^2+\phi'^2\right]\ .
\ee
By canonically normalising the field, i.e. by defining 
\be
\varphi\equiv\sqrt{\frac{\alpha^3}{\beta}}\phi\ ,
\ee
we have
\be
A=\frac{1}{2}\int dz dt\left[-\dot\varphi^2+\varphi'^2+V(z)\varphi^2\right]\ ,
\ee
where ${}'\equiv\partial_z$ and $V$ is a complicated function of $z$ which is not very enlightening to see. Recalling however that the system is strongly coupled at $y=y_c$, it is clear that the potential $V$ must have a divergence there. In figure \ref{divV} this divergence is shown explicitly (numerically). The reason the divergence is negative is simply a manifestation of the strong coupling. The divergence at large $y$ is instead the usual AdS barrier at spatial infinity.

We now search for mode solutions
\be
\varphi_\omega(t,z)=e^{i\omega t}h_\omega(z)\ .
\ee
Requiring $h_\omega$ to be normalizable, i.e. $\int_{0}^\infty |h_\omega|^2=1$. We have the boundary conditions 
\be\label{reg}
\lim_{z\rightarrow z(y_c),\infty}h_\omega=0\ .
\ee
The energy density of the system is the Hamiltonian. Each modes will then carry an energy density 
\be
{\cal E}_\omega=\frac{1}{2}V(z)|\varphi_\omega|^2+{\rm derivative\ contributions}\ .
\ee

\paragraph{Real frequencies: $\omega^2>0$}

In this case, the behaviour of the functions $h_\omega$ is very simple to guess. Defining $z_1$ and $z_2$ the two intersections of $|\omega^2|=|V|$, as in figure \ref{divV}, we have that for $z>z_2$ $h_\omega$ exponentially falls off. For $z_1<z<z_2$, by using the WKB approximation (see e.g. \cite{Davydov}), we instead have 
\be
h_\omega(z)\approx \frac{1}{\sqrt{z_2-z_1}}e^{i\omega z}
\ee
Finally for $z<z_1$, again using the WKB approximation and recalling that $\lim_{z\rightarrow z(y_c)}V=-\infty$, $h_\omega$ falls quickly off as (in $y$ coordinates, falls off exponentially)
\be\label{fall}
h_\omega(z)\propto \frac{1}{\left[\omega^2-V\right]^{1/4}}\ .
\ee
Note that the divergence of the potential acts as a rigid wall at the point $z=z(y_c)$.

Again, within the context of WKB, the frequencies will be thus quantised via the Born-Sommerfeld prescription\footnote{We have neglected the contribution coming from $z<z_1$.}
\be\label{phase}
\int_{z_1}^{z_2}\sqrt{\omega^2-V}=n\pi+k\ ,
\ee
where $k$ is a constant phase of the wave function and $n$ are integers.

Eq.\eqref{phase} approximately implies, as an order of magnitude estimate, 
\be
\Delta\omega\equiv\omega_i-\omega_{i-1}\approx \frac{1}{z_2-z_1}\ .
\ee
Considering a small $\lambda$, it is a straightforward exercise to find\footnote{Note that $z_2-z_1$ is also a function of $\omega$ and goes as $\sim\ln (\lambda/(R\omega))$, for all interesting physical values of $\omega$. However, for semiclassical values of $\omega R$, the $\ln\lambda$ dominates.}
\be
z_2-z_1\simeq -\frac{R}{\sqrt{4 G M}}\ln\lambda=\frac{R}{\sqrt{4 G M}}S\ .
\ee
Thus, the continuum Hawking spectrum is corrected by ${\cal O}(1/S)$ and becomes discrete, as suggested in \cite{dvali} and as required by the firewall arguments \cite{apologia}. 

In other words, a distant observer will declare the ``quantum corrected" Hawking spectrum to be unitary. However, the same observer, would also declare that the energy density close to the horizon diverges negatively as 
\be
{\cal E_\omega}\Big|_{z\rightarrow z(y_c)}\approx -\sqrt{|V|}\Big|_{z\rightarrow z(y_c)}\ ,
\ee 
since close to the pole of the potential (by using again WKB) $h_\omega\propto |V|^{-1/4}$. This matches our expectations, i.e. that the firewall is related to a divergent negative energy density close to the horizon, as found before in the Rindler case.

In the following, we shall discuss another source of firewalls due to the existence of imaginary frequencies. We will however see that, even in this extreme case, the firewall is only localised within a tiny reason close to the horizon, thus once again, only short wavelengths can probe it, as suggested in \cite{bousso}.

\paragraph{Imaginary frequencies: $\omega^2<0$}

Since the negative divergence of the potential is at $y=y_c$, imaginary frequencies are allowed. Those frequencies lead to an exponentially (in time) divergent energy density, which adds to the localised divergence of the energy density discussed above. However, even for imaginary frequencies, this divergent energy density is only localised within a tiny region close to the horizon. The reason is that, requiring regularity at infinity, the wave function $h_\omega$ exponentially decays in $z$. Of course though, being unbounded, the states with imaginary frequencies cannot be associated to any quantum states. Thus, a far observer might discard them claiming that some short scale physics (some UV completion) should project out those imaginary frequencies as usually happens in presence of singularities. Nevertheless, even in this case, the firewall associated to real frequencies (i.e. to genuine quantum states) would not be removed.

\paragraph{The size of the firewall}

To give an estimate of the size of the firewall, we can bound it by considering the point in which the potential passes through zero. Obviously this point will be an ${\cal O}(e^{-S})$ distance from the horizon \cite{bousso}. An explicit calculation indeed finds
$y_{\rm 0}\approx \frac{R}{\sqrt{80 G M}}\lambda$, i.e. the firewall is localised within a linear volume $\ell_{\rm firewall}\lesssim R e^{-S}$ (neglecting order one coefficients). Thus, as it is expected, the firewall disappears for infinite entropy.

What we have discovered in this section is that an asymptotic observer, believing that unitarity is restored by $e^{-S}$ ``corrections" to the Hawking spectrum, would infer that a free falling observer sees a firewall at a distance $\ell_{\rm firewall}$ to the horizon. Of course, as we already know, this hypothesis is only due to an inconsistent truncation of non-perturbative effects at order ${\cal O}(e^{-S})$.

\section{Conclusion}\label{concl}

In the context of quantum field theory, correlation functions of quantum operators are usually dominated by Lorentzian saddle points of the path integral. However gravity turns out to be a very special case that contradicts this expectation. 

Correlation functions calculated on a black hole spacetime of, e.g., scalar operators, eventually exponentially decay in time. This same fact signals the break-down of predictability, or loss of unitarity, of the back hole systems. For an asymptotically AdS black hole, in the AdS/CFT framework,  however, unitarity cannot be lost since the boundary CFT {\it is} unitary. Unitarity can then only be retrieved if the path integral for gravity is {\it not} dominated, at least after long time, by Lorentzian saddle points. This is clear. As the CFT is supposed to be dual to the bulk spacetime, it must have also access to the inside horizon. By a judicious calculation of the correlation functions one finds, at least in the two and three dimensional toy models, that Euclidean saddle points are indeed as relevant as the Lorentzian ones \cite{maldacena,Germani:2013sra}. However, as it has been proven by \cite{Kleban:2004rx}, even the contribution of Euclidean saddle points cannot solve the problem of unitarity for black holes. One therefore expects some non-perturbative effect to play an important role.

By assuming that the black hole geometry is only corrected by some operators of order ${\cal O}(e^{-2S})$ it has been proven that the energy momentum tensor of in-falling observer must necessarily diverge at the horizon \cite{amps}. This is the so-called firewall paradox. It is a paradox because one would immediately conclude that the assumption of small corrections to the black hole geometry fails right at the horizon, due to a large amount of energy stored there. 

Assuming the existence of a semiclassical black hole at early times, we showed that corrections to the black hole metric that solves the information paradox are instead of order ${\cal O}(g^{tt}e^{-2S})$. While being small far away from the horizon, they are dominant at the horizon. These (re-summed) corrections all together avoid the firewall and the information paradoxes. 
However, insisting on expanding the non-perturbative corrections in terms of the naive parameter ${\cal O}(e^{-2S})$, we showed that a distant observer (while observing a discrete Hawking spectrum) would infer the existence of a fictitious firewall at the horizon, precisely as discussed in \cite{apologia}. Thus, the message here is that {\it a firewall simply signals the limitations of the expansion in terms of saddle point approximations}.

Our findings address the problem of the {\it very long time} behaviour of correlation functions on a quantum ``black hole" spacetime. What we have proved is that, from Page time onwards, quantum corrections to the semiclassical solution must be non-perturbative in order to avoid the firewall paradox. In other words, at the Page time, no matter how big the black hole is, semiclassical physics breaks down. Then by continuity, early time corrections to the Hawking radiation {\it cannot} be exponentially suppressed by the entropy (unless one accepts the firewall), as usually naively assumed. It is very interesting to see that this independent conclusion is quite similar to the suggestions of \cite{violation}, where inverse power law corrections in entropy to the {\it early} Hawking radiation are proposed. In that scenario, the Page time is, as in our case, the moment in which any perturbative expansion to the semiclassical ``black hole" breaks down.

%

\vspace{0.5 cm}
\centerline{\bf Acknowledgements}
\noindent

We thank G. Dvali for comments on the first version of our paper and A. Kidambi for proof reading. DS also thanks D.~N.~Kabat and S.~N.~Solodukhin for comments. CG was supported by the Humboldt foundation during the first part of the project and wish to thank Ludwig-Maximilians-University for hospitality. DS is supported by the ERC Self-completion grant.
%

\providecommand{\href}[2]{#2}\begingroup\raggedright

\end{document}